\documentclass[12pt,a4paper]{article}
\usepackage[utf8]{inputenc}
\usepackage{amsmath,amssymb}
\usepackage{graphicx}
\usepackage{hyperref}
\usepackage{geometry}
\usepackage{soul}
\usepackage{color}
\usepackage{caption}
\usepackage{subcaption}
\usepackage{placeins}
\usepackage{authblk}

\title{An evacuation simulator for pedestrian dynamics based on the Social Force Model}
\author[1,2]{Juli\'an L\'opez}
\author[1]{Virginia Mazzone}
\author[2,3]{M. Leticia {Rubio Puzzo}}
\author[1,2]{Juan Cruz Moreno \thanks{juancruz.moreno@unq.edu.ar}}
\affil[1]{Laboratorio SiCoBioNa, Universidad Nacional de Quilmes, Argentina}
\affil[2]{Consejo Nacional de Investigaciones Cient\'ificas y T\'ecnicas (CONICET), Argentina}
\affil[3]{Instituto de F\'isica de L\'iquidos y Sistemas Biol\'ogicos, Universidad Nacional de La Plata, Argentina}

\date{\today}

\begin{document}

\maketitle

\begin{abstract}
The evacuation of pedestrians from enclosed spaces represents a key problem in safety engineering and infrastructure design. Analyzing the collective dynamics that emerge during evacuation processes requires simulation tools capable of capturing individual interactions and spatial constraints realistically.

In this work, we present \textit{SiCoBioNa}, an open-source evacuation simulator based on the Social Force Model (SFM). The software provides an intuitive graphical interface that allows users to configure pedestrian properties, spatial geometries, and initial conditions without requiring prior expertise in numerical modeling techniques. The SFM framework enables the representation of goal-oriented motion, interpersonal interactions, and interactions with fixed obstacles.

The simulator generates both quantitative data and visual outputs, facilitating the analysis of evacuation dynamics and the evaluation of different spatial configurations. Due to its modular and extensible design, \textit{SiCoBioNa} serves as a reproducible research tool for studies on pedestrian dynamics 
providing practical support for evacuation planning.
\end{abstract}


\section{Introduction}

The evacuation of pedestrians from enclosed or semi-enclosed environments such as airports, shopping centers, schools, or stadiums constitutes a relevant problem in safety engineering and architectural design. Evacuation processes may occur under normal operating conditions or during emergencies triggered by fire, panic, or external threats. In both cases, pedestrian motion through corridors, exits, stairways, and bottleneck regions results in collective dynamics where individual interactions, physical obstacles, and stress-induced behaviors can generate congestion, flow interruptions, or crowd disasters. Empirical observations and numerical studies have demonstrated that understanding these dynamics is essential for evacuation planning, exit optimization, and reducing evacuation time, highlighting the importance of parameters such as pedestrian speed, density, spatial geometry, and the emergence of effects such as the faster-is-slower phenomenon \cite{HelbingMolnar1995,Helbing2000,Xu2023,Dukker2025}.

Several physical and behavioral models have been proposed in the literature to describe and predict collective pedestrian motion based on a limited set of variables \cite{Zheng2009}. However, the availability of evacuation simulators that implement these models in a simple, intuitive, and visually accessible manner remains crucial for both scientific analysis and practical applications. Such tools allow researchers and decision-makers to virtually explore different spatial configurations, population densities, initial conditions, and behavioral assumptions without the costs and risks associated with real-world experiments. 
In addition, simulators—based on validated physical and behavioral models—facilitate the comparison of evacuation strategies, estimation of evacuation times, identification of bottlenecks, and evaluation of mitigation measures such as signage or structural modifications.

In this work, we present \textit{SiCoBioNa}, an evacuation simulator designed to model pedestrian motion during evacuation scenarios. The main objective of the simulator is to provide realistic numerical experiments that enhance the analysis and understanding of collective pedestrian behavior across various geometries and conditions. By adjusting a limited number of parameters, such as desired speed or behavioral response, the simulator can reproduce typical evacuation scenarios ranging from orderly pedestrian flow to panic-driven crowd motion.

The \textit{SiCoBioNa} simulator has been developed in Python (Python~3) \cite{van1995python,Python} and is distributed as open-source software, with its source code available for free. To enhance usability, a graphical user interface has been implemented, allowing users to configure pedestrian properties, environmental geometry, and simulation parameters without requiring prior knowledge of numerical modeling or programming. Additionally, the simulator incorporates automatic consistency checks to detect configuration errors and ensure the validity of the simulated scenarios.

In the version presented here, \textit{SiCoBioNa} simulates evacuation processes based on the Social Force Model (SFM) \cite{HelbingMolnar1995,Helbing2000}. The SFM has been extensively applied to the study of pedestrian dynamics and evacuation strategies in public spaces \cite{SimulationLargeScale2019,Zhang2025}. Its success has motivated numerous studies aimed at improving realism by incorporating additional interaction mechanisms. In the present work, the simulator relies exclusively on the standard SFM formulation, which provides a well-established physical framework for describing pedestrian motion.

The simulator allows the modeling of evacuation scenarios involving a main corridor connected to multiple rooms, where pedestrians move toward an exit located at either end of the corridor. The dimensions of the corridor and rooms, as well as the number of pedestrians initially located in each area, can be freely specified.

This paper is organized as follows: Section \ref{models} provides a concise overview of the Social Force Model underlying the simulator. Section \ref{simulator} describes the structure and workflow of the \textit{SiCoBioNa} simulator. Finally, concluding remarks are discussed in Section \ref{conclusions}.

\section{Social Force Model (SFM)}
\label{models}

Although the main focus of this work lies on the evacuation simulator rather than on the detailed analysis of interaction models, this section provides a brief overview of the essential features of the Social Force Model (SFM). Readers interested in a comprehensive discussion of the model are referred to the original works \cite{HelbingMolnar1995,Helbing2000}.

The SFM is one of the most widely used models for pedestrian dynamics, where each pedestrian is represented as a point-like particle subjected to social forces. These forces account for three main contributions: (i) the tendency of pedestrians to move toward a desired target, such as an exit, (ii) interactions with other pedestrians aimed at avoiding collisions, and (iii) interactions with fixed obstacles such as walls or furniture. The model has proven effective in describing pedestrian motion in high-density scenarios and capturing the emergence of collective effects \cite{Helbing2009,Yu2005,Guo2013}.

In the SFM, pedestrians are modeled as particles with mass and circular cross-section, with their orientation defined such that the forward direction is aligned with the velocity vector. Denoting by $\mathbf{x}_i$ the position of the $i$-th pedestrian with mass $m_i$, the equations of motion read
\begin{align}
    \dot{\mathbf{x}}_i &= \mathbf{v}_i, \label{eq:sfm1} \\
    m_i \dot{\mathbf{v}}_i &= \mathbf{f}_i^0 + \sum_{j (\neq i)} \mathbf{f}_{ij} + \sum_{W} \mathbf{f}_{iW},
    \label{eq:sfm2}
\end{align}
where $\mathbf{f}_i^0$ is the desired force, $\mathbf{f}_{ij}$ represents pedestrian--pedestrian interactions, and $\mathbf{f}_{iW}$ accounts for interactions with walls or obstacles.

The desired force models the acceleration required for a pedestrian to reach a desired speed $v_i^0$ in the direction $\mathbf{e}_i^0$ over a characteristic relaxation time $\tau_i$,
\begin{equation}
    \mathbf{f}_i^0 = m_i \frac{v_i^0 \mathbf{e}_i^0 - \mathbf{v}_i}{\tau_i}.
\end{equation}

Pedestrian--pedestrian interactions combine social repulsion and physical contact forces. Let $r_{ij} = r_i + r_j$ denote the sum of pedestrian radii, and $d_{ij} = |\mathbf{r}i - \mathbf{r}j|$ be the distance between centers, with $\mathbf{n}{ij}$ as the normalized vector from pedestrian $j$ to $i$, and $\mathbf{t}{ij}$ as the corresponding tangential direction. The interaction force is given by:

\begin{equation}
    \mathbf{f}_{ij} =
    \left[
    A_i \exp\left(\frac{r_{ij} - d_{ij}}{B_i}\right) + k g(r_{ij} - d_{ij})
    \right]\mathbf{n}_{ij}
    + \kappa g(r_{ij} - d_{ij}) \Delta \nu_{ji}^t \mathbf{t}_{ij},
\end{equation}
where $g(x)$ is zero for $x < 0$ and equal to $x$ otherwise, $\Delta \nu_{ji}^t$ denotes the tangential velocity difference, and $A_i$, $B_i$, $k$, and $\kappa$ are model parameters.

Interactions with walls or obstacles are modeled analogously:
\begin{equation}
    \mathbf{f}_{iW} =
    \left[
    A_i \exp\left(\frac{r_i - d_{iW}}{B_i}\right) + k g(r_i - d_{iW})
    \right]\mathbf{n}_{iW}
    + \kappa g(r_i - d_{iW}) (\mathbf{v}_i \cdot \mathbf{t}_{iW}) \mathbf{t}_{iW},
\end{equation}
where $d_{iW}$ denotes the distance to wall $W$.

In the present simulator, parameter values commonly adopted in the literature are used \cite{HelbingMolnar1995}: $A_i = 2 \times 10^3$~N, $B_i = 0.08$~m, $k = 1.2 \times 10^5$~kg~s$^{-2}$, $\kappa = 2.4 \times 10^5$~kg~m$^{-1}$~s$^{-1}$, and $\tau = 0.5$~s.

\section{The SiCoBioNa Simulator}
\label{simulator}

\subsection{Simulator Structure}

\begin{figure}[h!]
\centering
\includegraphics[width=1.0\linewidth]{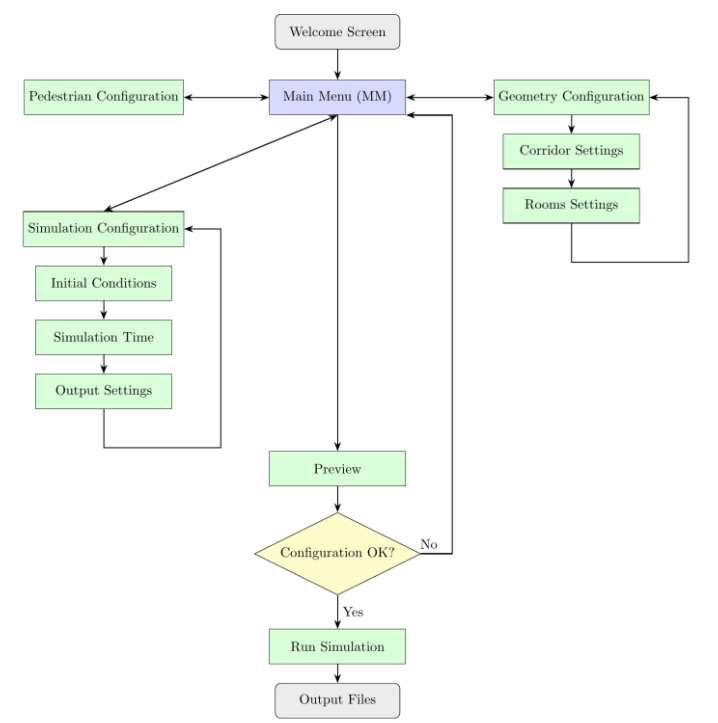}
\caption{Flowchart of the evacuation simulation software. The diagram illustrates the structured interaction between configuration modules, preview, validation, execution, and automatic generation of output files, preserving the original layout while avoiding graphical overlaps.}
\label{fig:software_flowchart_clean}
\end{figure}

The simulator workflow is designed to guide the user sequentially through the required steps to configure, visualize, and execute an evacuation simulation, ensuring an intuitive and structured user experience. The overall workflow is summarized in the flowchart shown in Fig.~\ref{fig:software_flowchart_clean}.


\begin{figure}[h!]
\centering
\includegraphics[width=0.5\linewidth]{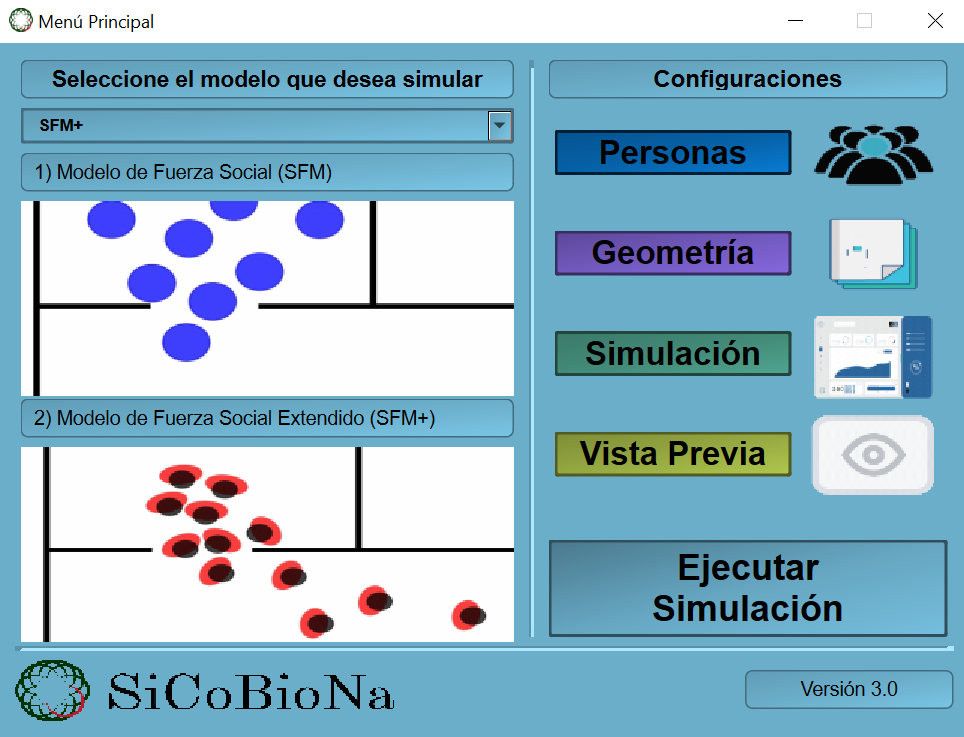}
\caption{Main menu of the SiCoBioNa simulator.}
\label{fig:menu}
\end{figure}

After launching the program and displaying a welcome screen, the user is directed to the main menu (Fig.~\ref{fig:menu}), from which all simulation configuration options can be accessed. To improve usability, the graphical interface incorporates visual cues and distinct color schemes for each configuration module.

The first required step is the selection of the physical model. In the current version, only the Social Force Model (SFM) is available; however, the software architecture allows for the straightforward incorporation of additional physical models in future releases. This choice conditions the available configuration options for pedestrian properties and simulation parameters.

\begin{figure}[htbp]
\centering
\setlength{\tabcolsep}{6pt}

\begin{tabular}{cc}

    \begin{subfigure}[b]{0.46\textwidth}
        \centering
        \includegraphics[height=0.27\textheight, keepaspectratio]{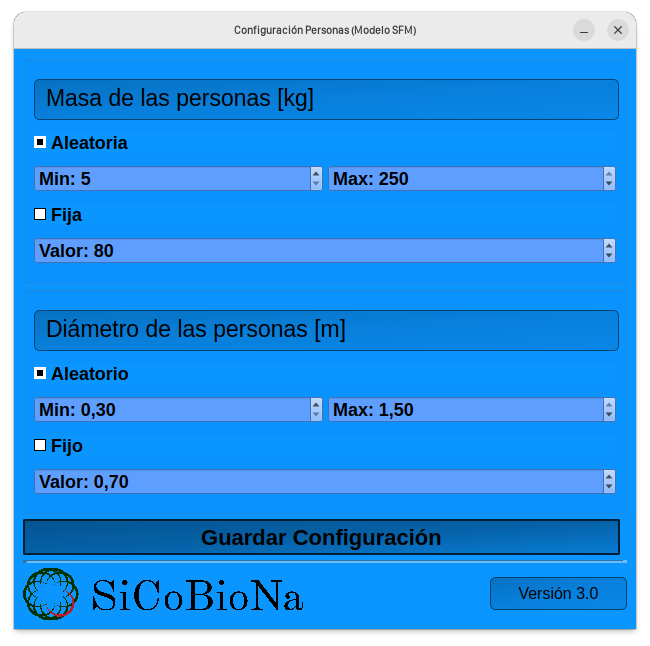}
        \caption{Pedestrian parameter configuration (SFM).}
    \end{subfigure} &
    \begin{subfigure}[b]{0.46\textwidth}
        \centering
        \includegraphics[height=0.27\textheight, keepaspectratio]{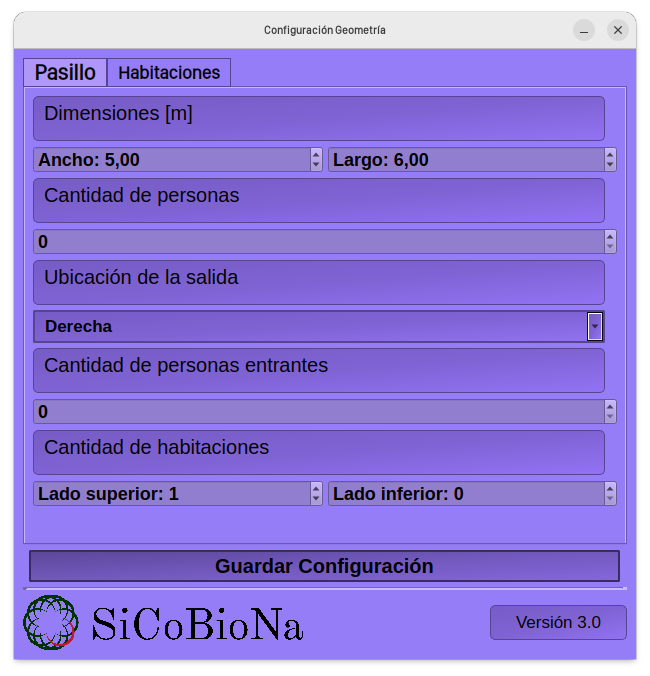}
        \caption{Corridor geometry configuration.}
    \end{subfigure}
    \\[8pt]

    \begin{subfigure}[b]{0.46\textwidth}
        \centering
        \includegraphics[height=0.27\textheight, keepaspectratio]{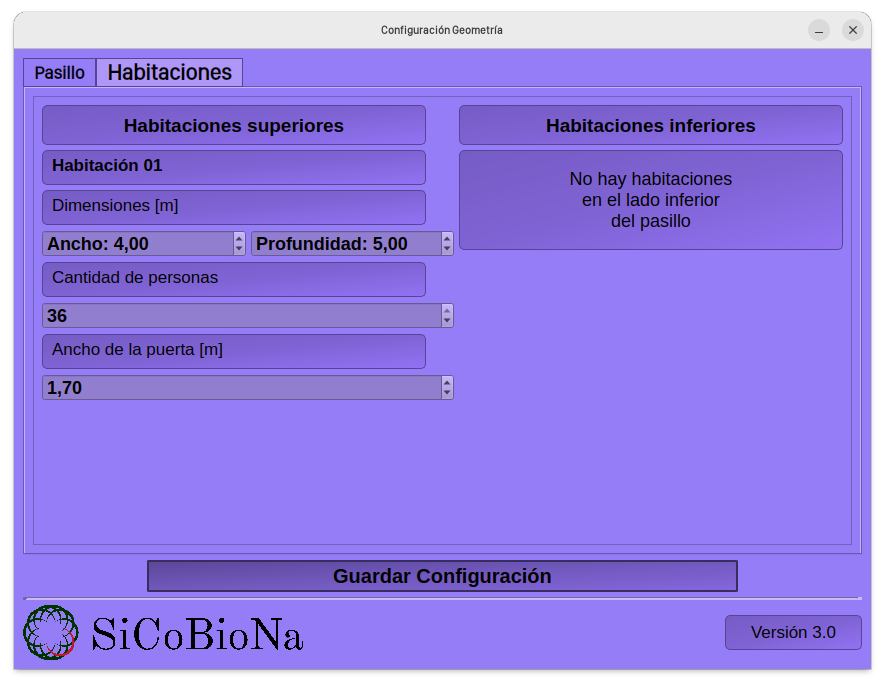}
        \caption{Room geometry configuration.}
    \end{subfigure} &
    \begin{subfigure}[b]{0.46\textwidth}
        \centering
        \includegraphics[height=0.27\textheight, keepaspectratio]{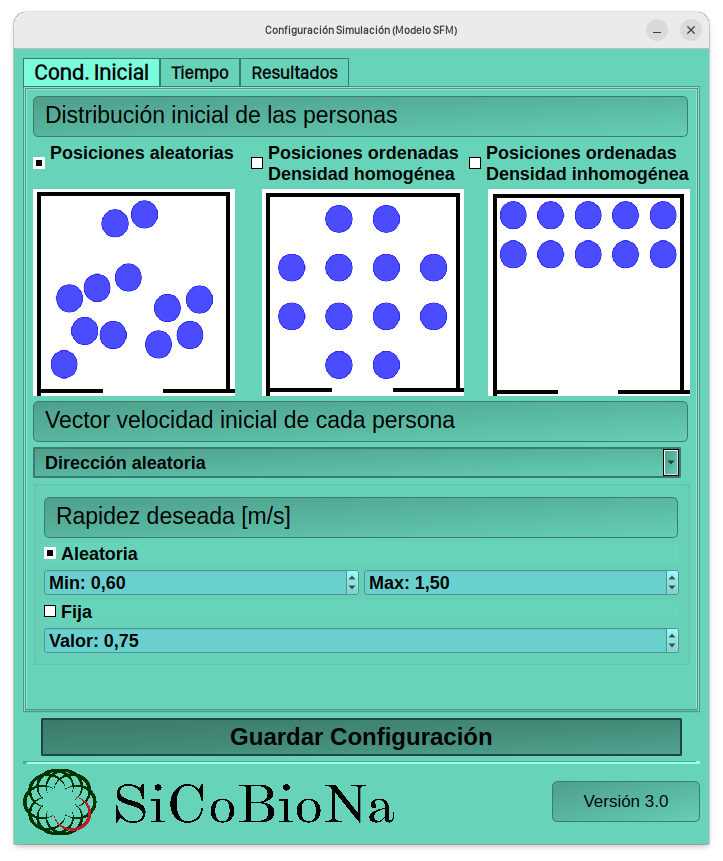}
        \caption{Initial condition configuration.}
    \end{subfigure}
    \\[8pt]

    \begin{subfigure}[b]{0.46\textwidth}
       \centering
        \includegraphics[height=0.15\textheight, keepaspectratio]{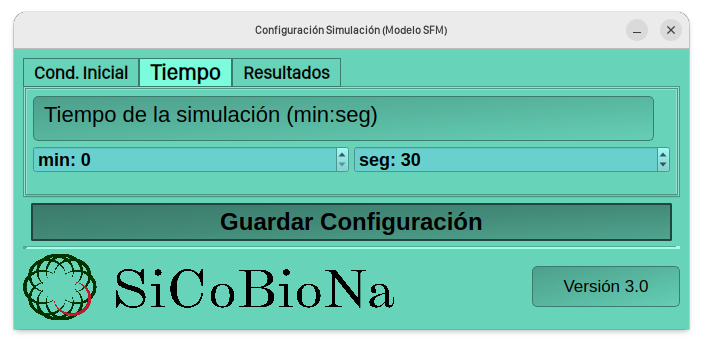}
        \caption{Simulation time settings.}
    \end{subfigure} &
    \begin{subfigure}[b]{0.46\textwidth}
       \centering
        \includegraphics[height=0.27\textheight, keepaspectratio]{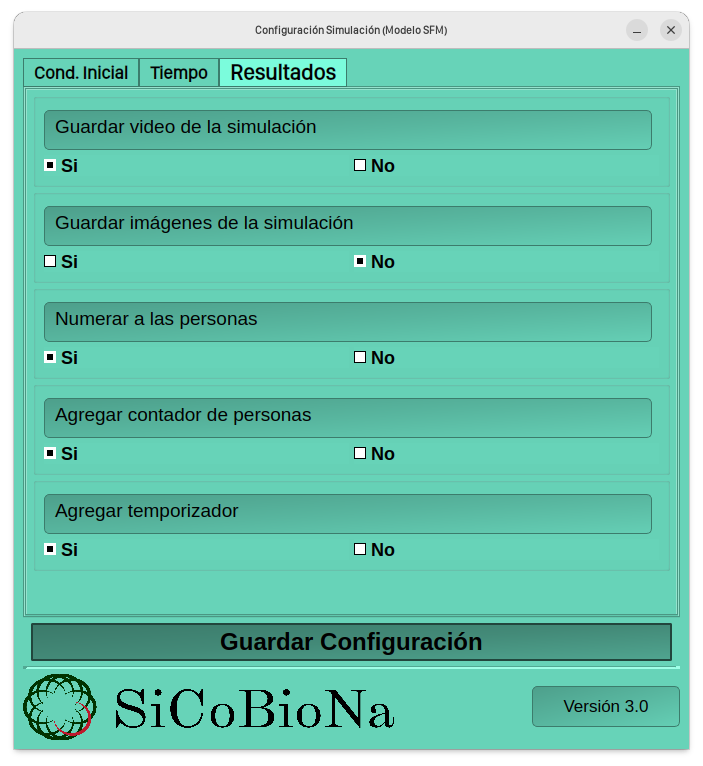}
        \caption{Output and visualization settings.} 
    \end{subfigure}

\end{tabular}

\caption{Set of simulator screen interfaces for configuring evacuation.}
\label{fig:simulador_SFM}
\end{figure}

The next recommended step is the configuration of pedestrian parameters. After selecting this option, a dedicated window is displayed (Fig.~\ref{fig:simulador_SFM}(a)), where pedestrians' masses can be defined either as fixed values or as random values within a user-specified range. In the same window, the pedestrian's diameter must be specified, recalling that pedestrians are modeled as circular particles in the SFM. Once the configuration is saved, the simulator automatically returns to the main menu.

Subsequently, the evacuation geometry must be defined. In all cases, evacuation occurs through a corridor leading to an exit located either on the left or the right side (Fig.~\ref{fig:simulador_SFM}(b)). The user can specify the number of pedestrians initially located in the corridor, as well as the number of pedestrians entering the corridor during the simulation. The number and position of rooms connected to the corridor (upper or lower side) can also be configured. The room configuration panel (Fig.~\ref{fig:simulador_SFM}(c)) allows the user to define the dimensions of the rooms, the initial number of pedestrians in each room, and the width of the exit doors. Each room can be configured independently.

The simulation configuration module allows the user to define the initial spatial distribution of pedestrians using three options (Fig.~\ref{fig:simulador_SFM}(d)): (i) random positions, (ii) ordered positions with homogeneous density, and (iii) ordered positions with inhomogeneous density while maintaining equidistant rows. Initial velocity directions can be defined as either random or oriented toward the exit. The desired speed can be set as fixed or randomly distributed within a prescribed range. 
A dedicated panel (Fig.~\ref{fig:simulador_SFM}(e)) is used to configure the simulation time. 
The output configuration panel (Fig.~\ref{fig:simulador_SFM}(f)) allows the user to select whether to generate simulation videos, image snapshots, pedestrian identifiers, counters, and on-screen timers. 
After saving the configuration, the simulator returns to the main menu.

\FloatBarrier

\subsection{Simulation Execution and Output}

\begin{figure}[htbp]
\centering
\setlength{\tabcolsep}{6pt}

\begin{tabular}{cc}

    \begin{subfigure}[b]{0.46\textwidth}
        \centering
        \includegraphics[height=0.27\textheight, keepaspectratio]{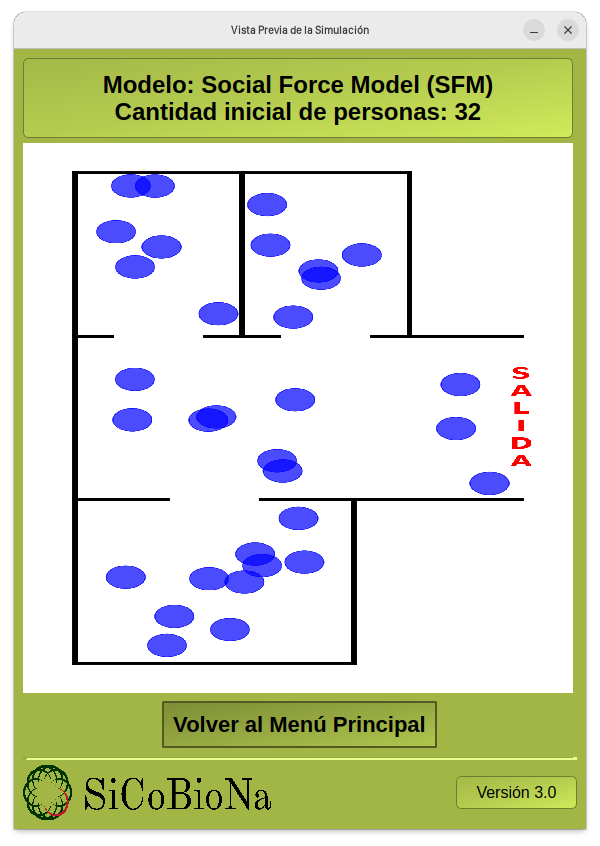}
        \caption{Preview of the initial simulation configuration.}
    \end{subfigure} &
    \begin{subfigure}[b]{0.46\textwidth}
        \centering
        \includegraphics[height=0.27\textheight, keepaspectratio]{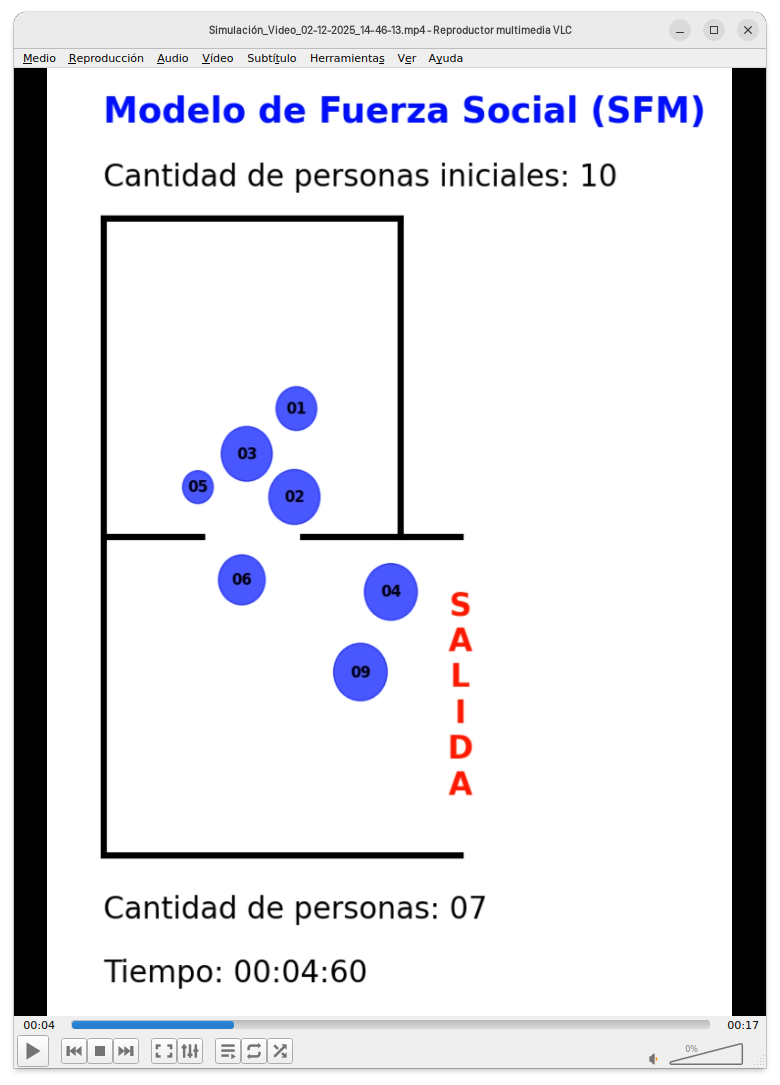}
        \caption{Representative frame from the simulation video.}
    \end{subfigure}
    \\[8pt]
\end{tabular}
    \caption{(a) Preview window showing the initial conditions of the evacuation scenario, including room layout and pedestrian distribution. (b) Screenshot of the simulation video, where circle sizes are consistent with pedestrian dimensions.}

\label{fig:VistaPrevia}
\end{figure}

Prior to execution, a preview window can be accessed from the main menu, providing a visual representation of the selected geometry and the initial number of pedestrians (Fig.~\ref{fig:VistaPrevia}(a)). This step allows the user to verify and, if necessary, modify the configuration before running the simulation.

Once execution is initiated from the main menu, the simulation runs in the background. Progress windows inform the user about the simulation status and the generation of images or video outputs, if selected.

\begin{figure}[htbp]
\centering
\includegraphics[width=0.65\linewidth]{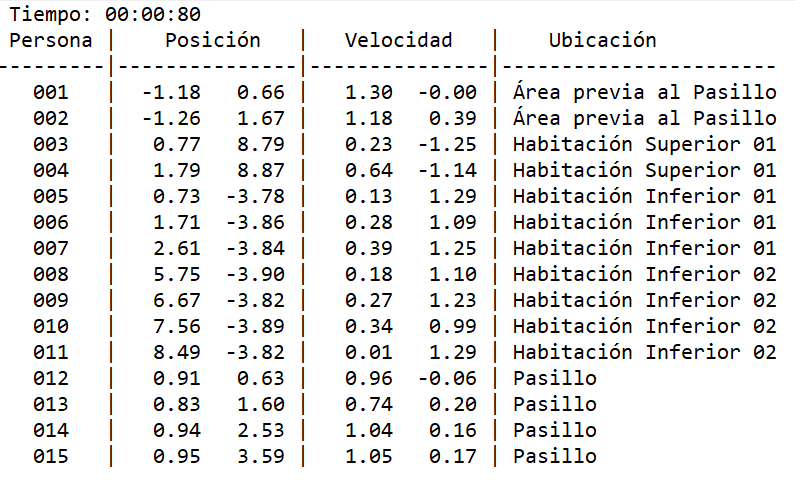}
\caption{Example of a simulation output data file generated using the Social Force Model.}
\label{fig:TXT}
\end{figure}

As previously described, the simulation output consists of a sequence of numbered text files. Each file's columns correspond to pedestrian ID, position coordinates, velocity components, and spatial location (Fig.~\ref{fig:TXT}). These files are generated at regular time intervals, every 10s of simulated time. They are subsequently used to produce a video file in \texttt{.mp4} format.

\subsubsection{Error Detection}

The simulator includes an alert system based on pop-up windows that notify the user of configuration errors prior to execution. The most common issues are related to incompatible geometric configurations. For example, a corridor's length may be insufficient for the selected room dimensions. These validation mechanisms prevent inconsistent simulations and ensure the physical coherence of the configured scenarios.

\FloatBarrier

\section{Comments and Conclusions}
\label{conclusions}

In this work, we have presented \textit{SiCoBioNa}, an open-source evacuation simulator based on the Social Force Model. The simulator enables realistic numerical experiments aimed at analyzing pedestrian dynamics during evacuation processes.

The availability of both numerical output data and visual representations allows for detailed studies of evacuation dynamics and facilitates the investigation of the underlying physical mechanisms described by the SFM.

The modular design of the simulator ensures extensibility, allowing for future incorporation of alternative interaction models or more complex geometrical configurations. 
This flexibility makes \textit{SiCoBioNa} suitable for both research-oriented and applied studies related to safety assessment and infrastructure design.

Finally, the simulator is publicly available through the \textit{SiCoBioNa Laboratory} website ($http://sicobiona.web.unq.edu.ar/proyectos/simulador$-$sicobiona/$). While the current version is distributed in Spanish, an English-language version will be released in the near future.

\section*{Acknowledgements}
This work was supported by the National Council for Scientific and Technical Research (CONICET), the National University of Quilmes, and the National University of La Plata (Project 11/X973). The authors J. López, V. Mazzone and J. C. Moreno express their special gratitude to the members of the SiCoBioNa Laboratory, particularly F. Alvira and A. Otero.

\bibliographystyle{unsrt}
\bibliography{bibPedestrian}

\end{document}